\documentclass{aa}  
\usepackage{graphicx}
\usepackage{txfonts}
\usepackage{longtable}

\newcommand{\mj}{$M_\mathrm{J} \ $}

\begin{document} 
   \title{High contrast imaging at 10 microns, a search for exoplanets around:  Eps Indi A, Eps Eri, Tau Ceti, Sirius A and Sirius B}
   
   \author{P. Pathak\inst{1} \and D. J. M. Petit dit de la Roche\inst{1} \and M. Kasper\inst{1} \and M. Sterzik\inst{1} \and O. Absil\inst{2}\fnmsep\thanks{F.R.S.-FNRS Research Associate}\and A. Boehle\inst{3} \and F. Feng\inst{8,9} \and V. D. Ivanov\inst{1} \and M. Janson\inst{5} \and H.R.A. Jones\inst{4} \and A. Kaufer\inst{1} \and H.-U. Käufl\inst{1} \and A.-L. Maire\inst{2} \and M. Meyer\inst{7} \and E. Pantin\inst{6} \and R. Siebenmorgen\inst{1} \and M. E. van den Ancker\inst{1} \and G. Viswanath\inst{5}
   }
   \institute{European Southern Observatory, Karl-Schwarzschild-Str. 2, 85748 Garching bei München, Germany \\
   \email{ppathak@eso.org}
    \and Space sciences, Technologies and Astrophysics Research (STAR) Institute, Universit\'e de Li\`ege, 19c All\'ee du Six Ao\^ut, 4000 Li\`ege, Belgium
    \and Institut für Teilchen- und Astrophysik ETH Zurich, Switzerland
    \and Center for Astronomy Research, University of Hertfordshire, Hatfield, United Kingdom
    \and Stockholm University, Stockholm Observatory, Sweden
    \and AIM, CEA, CNRS, Université Paris-Saclay, Université Paris Diderot, Sorbonne Paris Cité, F-91191 Gif-sur-Yvette, France
    \and Astronomy Department, University of Michigan, Ann Arbor, MI 48109, USA
    \and Tsung-Dao Lee Institute, Shanghai Jiao Tong University, 800 Dongchuan Road, Shanghai 200240, People's Republic of China
    \and Department of Astronomy, School of Physics and Astronomy, Shanghai Jiao Tong University, 800 Dongchuan Road, Shanghai 200240, People's Republic of China
   }
   \date{Received xx, 2020; accepted xx, 2020}

  \abstract
   {The direct imaging of rocky exoplanets is one of the major science goals for upcoming large telescopes. The contrast requirement for imaging such planets is challenging. However, the mid-IR (InfraRed) regime provides the optimum contrast to directly detect the thermal signatures of exoplanets in our solar neighbourhood.}  
   {We aim to exploit novel fast chopping techniques newly developed for astronomy with the aid of adaptive optics to look for thermal signatures of exoplanets around bright stars in the solar neighbourhood.}
   {We use the upgraded VISIR (Very Large Telescope Imager and Spectrometer for the mid-InfraRed) instrument with high contrast imaging (HCI) capability optimized for observations at 10~$\mu$m to look for exoplanets around five nearby ($d$ < 4 pc) stars. The instrument provides an improved signal-to-noise (S/N) by a factor of $\sim$4 in the N-band compared to standard VISIR for a given S/N and time.}
  {In this work we achieve a detection sensitivity of sub-mJy, which is sufficient to detect few Jupiter mass planets in nearby systems. Although no detections are made we achieve most sensitive limits within $<2''$ for all the observed targets compared to previous campaigns. For $\epsilon$ Indi A and $\epsilon$ Eri we achieve detection limits very close to the giant planets discovered by RV, with the limits on $\epsilon$ Indi A being the most sensitive to date. Our non-detection therefore supports an older age for $\epsilon$ Indi A. The results presented here show the promise for high contrast imaging and exoplanet detections in the mid-IR regime.}
  {}

\maketitle
%

\section{Introduction}
The direct imaging of habitable exoplanets is one of the key science goals of current and upcoming large telescopes~\citep[][]{meyer2018}. The field of high-contrast imaging (HCI), employing extreme adaptive optics (ExAO), coronagraphy and state-of the art post-processing techniques, has enabled direct imaging of young (up to about 30 Myr), several Jupiter-mass exoplanets with current $8-10~m$ class telescopes~\citep{morois2008,macintosh2015,keppler2018,chauvin2017,nowak2020}. Examples of current HCI instruments include the Spectro-Polarimetric High-contrast Exoplanet REsearch instrument (SPHERE), the Gemini Planet Imager (GPI) and Subaru Coronagraphic Extreme Adaptive Optics (SCExAO), all of which operate in the near-IR (1-2.5 $\mu$m) regime~\citep{sphere, gpi, scexao}. Compared to the near-IR, the mid-IR (8-13$\mu$m) is more sensitive to colder planets and allows to probe less massive planets or, for a given mass, one is able to search around older stars \citep{quanz2015}. This is because the planet to star flux contrast is more favourable in the mid-IR, where the thermal emission of the planet peaks in the Rayleigh-Jeans tail of target stars~\citep[][]{Baraffe2003,Sudarsky2003,Marley2007,Fortney2008,Spiegel2012}. The key downsides of mid-IR HCI are reduced spatial resolution due to the larger diffraction limit and large sky-background for ground-based observations. Therefore, the mid-IR is best suited to look for exoplanets around nearby stars. 

New Earths in the $\alpha$ Cen Region (NEAR) experiment was a collaboration between the Breakthrough Foundation and the European Southern Observatory (ESO). The project involved upgrading the existing VISIR (Very Large Telescope Imager and Spectrometer for the mid-InfraRed) instrument~\citep{lagage2004} at the VLT with adaptive optics (AO) using the deformable secondary mirror (DSM) installed at UT4~\citep{Arsenault2017}, and a high-performance annular groove phase mask (AGPM) coronagraph~\citep[][]{Mawet2005}. The aim of the NEAR experiment was to enable HCI capability in the astronomical N-band and to look for low mass exoplanets in the $\alpha$ Centauri binary system in a $100$~hr campaign~\citep[for details see][]{near2017, near2019}. The NEAR was able to reach sensitivity and contrast sufficient for detection of Neptune mass planets in the habitable zone of $\alpha$ Cen A, and a weak signal was found whose nature (e.g., planet, part of a zodiacal disk, image artefact) remains to be confirmed by follow-up observations~\citep[for details see][]{Wagner2021}.

In this work, we report the results of observations with NEAR to look for Jupiter size exoplanets around the nearest stars with spectral type earlier than M: $\epsilon$ Indi A, $\epsilon$ Eri, $\tau$ Ceti, Sirius A and Sirius B. 

In Sect.~\ref{sec:target} we briefly describe observed targets, in Sect.~\ref{sec:observations} and \ref{sec:results} we describe the data observation and reduction techniques. We discuss results in Sect.~\ref{sec:discussion} and conclude with Sect.~\ref{sec:conclusion}.

\section{Target description}\label{sec:target}
\subsection{$\mathrm{\epsilon}$ Indi A}
$\mathrm{\epsilon}$ Indi is a triple system at a distance of 3.6\,pc \citep{gaia3}. It consists of the primary K5V star $\mathrm{\epsilon}$ Indi A and a brown dwarf binary ($\mathrm{\epsilon}$ Indi Ba and Bb) on a wide orbit of 1459\,AU \citep{Scholz2003,McCaughrean2004}. Age estimates of $1.4^{+1}_{-0.5}$\,Gyr for $\epsilon$ Indi A were published based on chromospheric activity indicators such as the calcium R$_{HK}$ as a proxy for rotation \citep{lachaume1999}. Given the well-known relationships between rotation and age for FGK stars a similar value of $1.5$ Gyr was suggested by~\citet{Kasper2009}. However, \citet{Dieterich2018} suggest an older age based on UVW kinematics and cooling curves for the brown dwarf companions. Recently \citet{Feng2019} use extensive time-resolved spectra to estimate a rotation period for the star of 36 days, suggesting an age of $\sim$4\,Gyr based on the rotation-age calibration of \citet{Eker2015}. This older age also agrees with the 3.7-4.3\,Gyr estimated by \citet{King2010} for the brown dwarf binary from the dynamical system mass and the evolutionary models of \citet{Baraffe2003}. The higher end of the age range is therefore more likely.

\citet{Endl2002} first identified a long-period, low-amplitude radial velocity signal in $\mathrm{\epsilon}$ Indi A, which could be explained by a companion with P>20\,yr and a mass of at least 1.6\,\mj. This signal has been confirmed by \citet{Janson2009} and \citet{Zechmeister2013}, who also find that the binary brown dwarf companions $\mathrm{\epsilon}$ Indi Ba and Bb are too far away to induce the measured trend. \citet{Feng2019} combine radial velocity data with astrometry to confirm the existence of a $3.25^{+0.39}_{-0.65}$,\mj planet on an eccentric orbit with a period of $45.2^{+5.74}_{-4.77}$\,yr. In September 2019, at the time of our observations the separation of this planet from the host star is expected to be about 1.07$\arcsec$ (error bars are large due to poorly constrained orbital solution). 

\subsection{$\mathrm{\epsilon}$ Eri}
$\epsilon$ Eri is an adolescent K2V type dwarf star at a distance of $3.2$~pc~\citep{gaia3}. The age of the star has been estimated through various means and is generally thought to be around 0.4-0.9\,Gyr, with the higher end of the range being more likely \citep{Henry1996,Song2000,difolco2004,Mamajek2008}. $\epsilon$ Eri is surrounded by a narrow ring of debris located between 63 to 76 AU and a possible inner belt at 12 to 16 AU~\citep[see][for a full discussion of the disk structure]{mawet2019}. 

A companion to $\mathrm{\epsilon}$ Eri was first suggested by \citet{Walker1995} based on radial velocity data. \citet{Hatzes2000} argued that the most likely explanation for the observed decade-long radial velocity (RV) variations was the presence of a 1.5 \mj giant planet with a period P = 6.9 yr (3 AU orbit) and a high eccentricity (e=0.6). Very similar parameters for the planet $\epsilon$ Eri b were derived from a comprehensive set of RV as well HST astrometry data by~\citet{Benedict2006}. The most recent mass estimate to date was done by \citet{mawet2019} and combined 30 years of radial velocity data with deep direct imaging data in a Bayesian analysis to constrain the properties of the companion. They found a mass of $0.78^{+0.38}_{-0.12}$,~\mj at a separation of $3.48\pm0.07$\,AU and an eccentricity of $0.007$, which is lower than previous reported values. Direct imaging detections of $\epsilon$ Eri b in the L- and M-bands were attempted by \citet{Janson2008} and \citet{mawet2019} and yielded an upper mass limits of around 4~\mj (for an age of 320 Myr) and 2~\mj (for 400 Myr), respectively. A second companion was first suggested by \citet{Benedict2006} at 12-20\,AU based on radial velocity residuals, but this was not confirmed by \citet{mawet2019}. An alternative additional companion of 0.4-1.2\,\mj has been suggested at 48\,AU by \citet{Booth2017} to explain the shape of the outer dust belt, and \citet{mawet2019} also require an additional planet to stir this belt. \citet{Janson2015} are able to obtain sub-Jovian mass limits of 0.6-1\,\mj at these larger separations, but this cannot rule out the lower end of the proposed mass range, so while the companion has not been confirmed, it can also not been ruled out. 

\subsection{$\tau$ Ceti}
$\tau$ Ceti is a nearby (3.7\,pc), sun-like G8.5V star with an extended debris disk (5-55\,AU) that is more than 10 times as massive as the Kuiper belt \citep{gaia3,Gray2006,Greaves2004,MacGregor2016}. There is a large range in the literature for ages of $\tau$ Ceti: based on stellar activity the age is 5.8$\pm$2.9\,Gyr \citep{Mamajek2008}, but astroseismological and interferometric measurements suggest an age closer to 8-10\,Gyr \citep{difolco2004,Tang2011} and chemical composition measurements point towards an age of $7.63^{+0.87}_{-1.5}$\,Gyr \citep{Pagano2015}. 

\citet{Tuomi2013} discovered 5 Earth-like planets in radial velocity data with periods from 14 to 642 days (0.1-1.35\,AU). Two of these were confirmed by \citet{feng2017}, who discovered two further planets at periods of 20 and 49 days. They also suggest that the 14 day signal could be the result of stellar activity, rather than a planet and that candidates e, f and h might actually be too eccentric to be planets. \citet{Dietrich2020} use dynamical arguments to provide statistical evidence for the existence of the remaining three planets suggested by \citet{Tuomi2013} and one additional candidate that could be located in the habitable zone. All the previously mentioned candidates are expected to be super-Earths ($M\cdot sin(i)\approx 1-4M_\oplus$). However, \citet{Kervella2019} provided tentative evidence of a giant planet candidate (1-2\,\mj, 3-20\,AU) from Gaia data. Near infrared direct imaging has not been able to detect any of the planets, but has provided constraints of 10-20\,\mj at separations larger than 2$\arcsec$ (7AU) and 30-50\,\mj at 1$\arcsec$ \citep{boehle2019}. The candidate detected by \citet{Kervella2019} is the only one that is not too small and close to be detected with present-day mid-IR facilities.  

\subsection{Sirius A}
Sirius is a binary system at a distance of 2.7\,pc, consisting of an A1Vm star and a WD with an age of 225-250\,Myr \citep{gaia3,Liebert2005,bond2017}. While there are no known planets around Sirius A or B, there are mass limits on possible companions from previous imaging campaigns, covering wavelengths of roughly $0.5-5\,\mathrm{\mu}$m. The most sensitive limits on Sirius A exclude giant planets down to 11\,\mj at 0.5\,AU, 6-7\,\mj in the 1-2\,AU range and 4\,\mj at 10 AU~\citep{Hunziker2020,vigan2015,Thalmann2011,Bonnet2008}. In any case, the long-term orbital stability of planets around Sirius A or B would be impacted by the binarity of the system, which has a semi-major axis of about 20 AU and eccentricity 0.6.~\citet{bond2017} find that the longest periods for stable planetary orbits in the Sirius system are about 2.24 yr for a planet orbiting Sirius A, corresponding to a r = 2.2 AU circular orbit, and 1.79 yr for a planet orbiting Sirius B, corresponding to a r = 1.5 AU circular orbit. WDs are typically $10^3$ to $10^4$ times less luminous than their progenitor stars. Thus, it would be much easier to achieve the contrast required to detect a planetary companion~\citep{Burleigh2002}. This idea sparked direct imaging searches for planets around WDs~\citep{Graton2020,Friedrich2007}. While no planets have been found using direct imaging to date, a transiting planet was found around the white dwarf WD 1856+534~\citep{Vanderburg2020}, and an evaporating Neptune was proposed to explain the chemical fingerprints of hydrogen, oxygen and sulfur in the spectrum of WDJ0914+1914~\citep{gansicke2019}. 

\subsection{Sirius B}
For Sirius B, as explained above a planet in a stable orbit would have a period of 1.79 yr for a r = 1.5~AU. No limits have been placed on companions within 3$\arcsec$, although near infrared imaging by \citet{Bonnet2008} did set limits of 10-30\,\mj outside of that radius. Sirius B is the only target in the sample to have been previously imaged in the mid-IR at 8-10$\,\mu$m between 2003 and 2006, but no limits on companions were determined from this dataset \citep{skemer2011}. 

\begin{table*}[!htbp] 
	\begin{center}
		\begin{tabular}{lcccccccc} 
			\hline \hline%
			Target & Parallax & N-band (Jy) & Date & Observation & On-coronagraph & Seeing & 
			Temp ($^{\circ}$C)& PWV (mm)\\
            & (mas) & brightness & observed & time (hr) & time (hr) & (arcsec) &  & \\
			\hline
            $\epsilon$ Indi A & $274.84 $ & 4.5 & 2019-09-14 & 2.35 & 0.81 & 0.8-2.0 & 10-11 & 0.8-2.2\\
                     & & & 2019-09-15 & 4.23 & 1.47 & 0.5-1.0 & 12-13 & 0.4-0.9\\
                     & & & 2019-09-17 & 3.31 & 1.33 & 0.5-0.9 & 9-12 & 0.9-3.0\\
            \hline
            $\epsilon$ Eri & $310.58$ & 7.6 & 2019-09-15 & 1.61 & 0.61 & 1.0-1.8 & 11-12 & 0.4-0.7 \\
                     & & & 2019-09-17 & 2.63 & 1.01 & 0.5-0.9 & 9-10 & 0.8-1.4 \\
            \hline
            $\tau$ Ceti & $273.81$ & 7.5 & 2019-09-16 & 1.81 & 0.65 & 0.8-1.6 & 10-11 & 3.0-5.0 \\
            \hline
            Sirius A & $374.49$ & 118.8 & 2019-12-15 & 2.38 & 0.74 & 0.4-0.7 & 13-14 & 3.8-5.2 \\
        \hline
		\end{tabular}
	\end{center}
\caption{Observing parameters for all the targets under various conditions. All the observations were carried out using the NEAR N-band filter with a bandpass of $10-12.5~\mu$m. The N-band brightness reported in the table assumes the central wavelength of the NEAR filter of $11.25~\mu$m.}\tablefoot{The on-coronagraph time is $40\%$ of the selected frames due to the observing strategy based on chopping and nodding, for details see Sec.~\ref{sec:observations}. }
\label{tab:obs}
\end{table*}

\section{Observation and data reduction}\label{sec:observations}
All the observations employed a common observing strategy, including the use of AO, an AGPM coronagraph, chopping and nodding. The chopping was done with the DSM of the VLT. The chop throw of the DSM was $~4.5''$ and was performed at a speed of $8.33$~Hz to reduce the excess low frequency noise common to the mid-IR arrays~\citep[Si:As array,][]{Arrington1998}. The chopping subtracts two images taken at different position on the sky thereby leaving a positive (coronagraphic, on-axis) and a negative (off-axis) image of the source separated by the chop throw and removing most of the sky and instrumental background flux bias. However, the optical path is slightly different for both chopping positions, which leaves some small residuals. In principle, these could be removed by nodding~\citep[for details see][]{lagage2004}, but our data reduction is not affected by the small chopping residuals which are not point-like and do not degrade the point-source sensitivity. At 8.33 Hz chopping, each half-cycle is 60~ms consisting of $8\times6$~ms DITs (=$48$~ms) and $2\times6$~ms DITs (=$12$~ms) skipped during the chopping transition. The other half-cycle is taken with the star off the coronagraph. At each nodding position,  500 chopping frames half-cycles were recorded for a total observing time of $500\times60$~ms = $30$~seconds out of which $250\times48$~ms = 12 were spent with the target on the coronagraph, that is, the observing efficiency was 40$\%$. A summary of the observations and the atmospheric conditions affecting the sensitivity is outlined in the Table~\ref{tab:obs}. The atmospheric data were taken from Paranal Astronomical Site Monitoring. To center the targets on the coronagraph and to minimize the leakage, a dedicated correction using the science images with the aid of the QACITS algorithm was used~\citep[for details see][]{qacits2020}.  We derived a pixel scale of $45.25$ mas/pixel using the $\alpha$ Cen campaign data, utilizing the well known orbit of the binary from~\citet{Kervella2019}. In the next section we discuss the steps employed for the data reduction.

\begin{figure*}
	\centering
    \resizebox{\hsize}{!}{\includegraphics{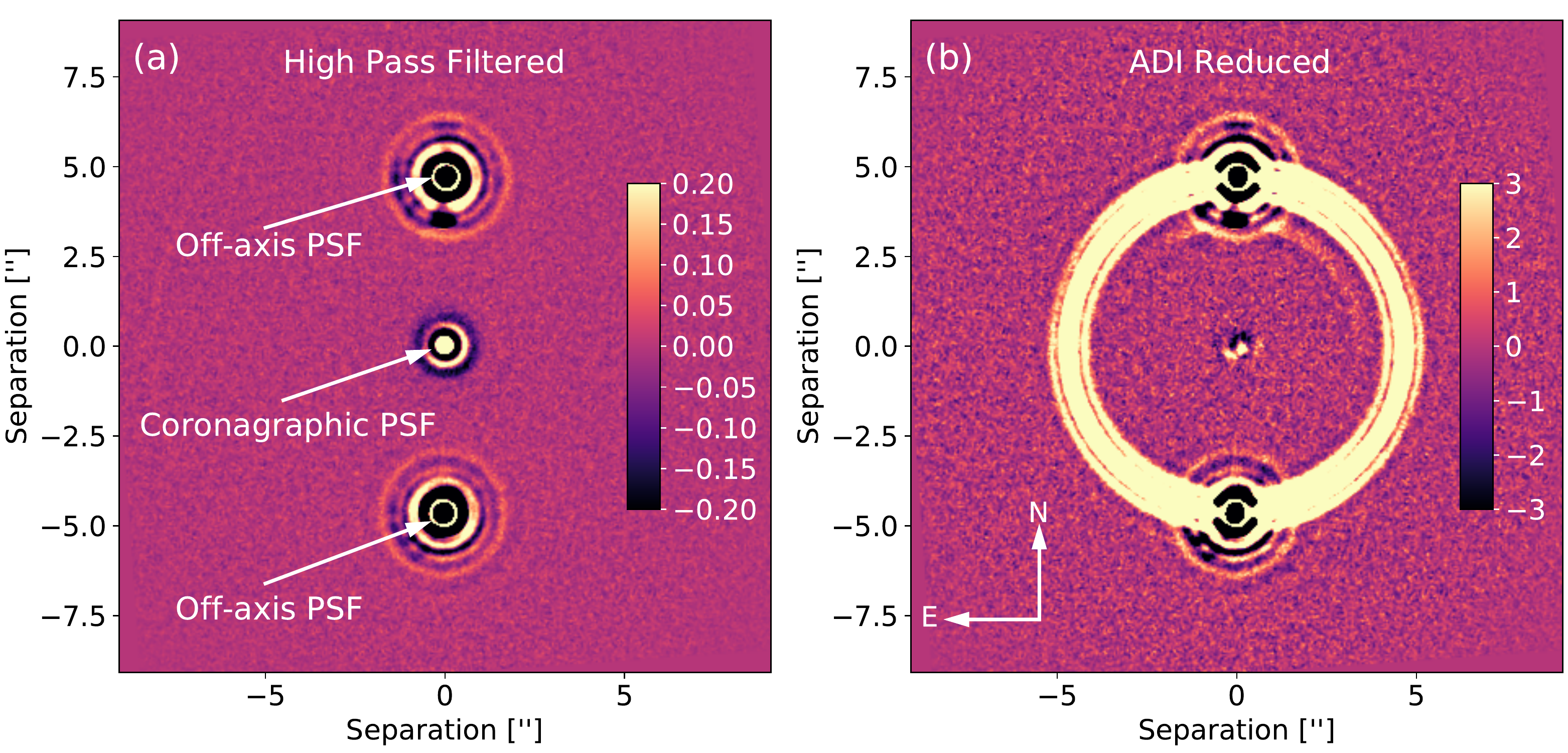}}
    \caption{\label{fig:data_red}(a) Final derotated image of the $\epsilon$ Indi A target with high-pass filter applied. (b) ADI processed image.}
\end{figure*}
\subsection{Data Reduction}
For all the targets a common data reduction strategy was followed, which included a chop subtraction of the off-axis from the on-axis source position frames adjacent in time. This provided images where the source was positive on the coronagraph and negative in the off-axis position. 

Three selection criteria were employed to identify inferior frames and remove them from the data analysis: AO correction (ratio of flux in an annulus of radii 6-12 pix to flux in an aperture of r<6 pix), coronagraphic leakage (flux in an aperture of 20 pix) and sky-background noise variance calculated using small regions of the non-chopped images. To further identify good, co-aligned frames, an additional parameter based on the positions of the off-axis PSFs was used. The employed selection criteria improved the overall sensitivity and contrast, and reduced false positives, especially at small projected separations. We lost about $14\%,~5\%,~10\%$ and $22\%$ of the observed frames for $\epsilon$ Indi A, $\epsilon$ Eri, $\tau$ Ceti and Sirius A respectively. 

Once good chopped frames were identified, they were binned by averaging 250 frames. The binning of the frames to an exposure time of 30 seconds ($250\times0.12$s) was chosen to be short enough to avoid smearing of potential companions by the field rotation, and long enough to provide a good sensitivity on binned frames and to reduce the data size for further post-processing. The averaged frames still show some smoothly varying structures (residual sky-background, left over after chopping), which was removed by applying a spatial high-pass filter. This filtering process creates a smoothed version of the image, by replacing each individual pixel by a median of $15\times15$ surrounding pixels, and then subtracts it from the image. The effect of such filtering on the point source signal was a flux reduction of less than $10\%$ and had a negligible effect on the noise variance. Figure \ref{fig:data_red} (a) shows a final averaged and derotated image for the target $\epsilon$ Indi A, with the coronagraphic PSF at the center and the off-axis PSF's representing both the nod positions.

For the final Angular Differential Imaging (ADI) analysis~\citep{Marois2006}, night by night data was processed using a global annular Principal component analysis (PCA) based algorithm. Specifically, for a given analysis frame, we identify all the frames obtained during that night which differ in field rotation angle by at least one PSF half width at half maximum (HWHM) at the smallest angular separation of interest. We set this separation to 8 pixels (or 360 mas) which corresponds to the 1st minimum of the VLT's N-band Airy pattern. For this set of calibration frames, first the mean of the set was subtracted from all individual frames. Then we select all pixels in an annular area with an inner radius of 8 pixels and an outer radius of 25 pix, corresponding to the 3rd minimum of the Airy pattern. As our images are usually very smooth and without residual speckle structure outside a radius of 25 pix (see~Figure~\ref{fig:data_red} a), we do not benefit from a larger outer radius. We perform the PCA analysis on these data, i.e., on the pixels in the annular area for the set of calibration frames arranged in a matrix of size $\#$pixels $\times$ $\#$frames. This yields the linear combinations of calibration frames (the principle components) which best reproduces the analysis frame. The optimization of the PCA parameters, such as inner and outer radius of the annulus and number of principal components, was done with artificial planet injection and recovery tests, to maximize the contrast sensitivity. We observed that 15 principal components yielded the best compromise between the reduction of PSF residuals and self-subtraction of artificial planets inserted into the data. A further analysis was performed by splitting the data into odd-even frames and dividing into different chunks, to see if any strong speckles remain for different analysis. This helped to identify suspected false positive detections such as a faint speckle visible just right of the coronagraph center in the Figure~\ref{fig:data_red} (b).

In the case of targets observed for more than one night, the final processed image was produced by weighting each night's combined image by the inverse of the background noise as measured in the combined frame of the night, i.e., by applying a noise weighted mean. This step was used to compensate for variable atmospheric conditions (ambient temperature, humidity and precipitable water vapor in the atmosphere), this helped to improve the final noise variance in the image~\citep[][]{turchi2020}. 

\begin{figure}
	\centering
    \resizebox{\hsize}{!}{\includegraphics{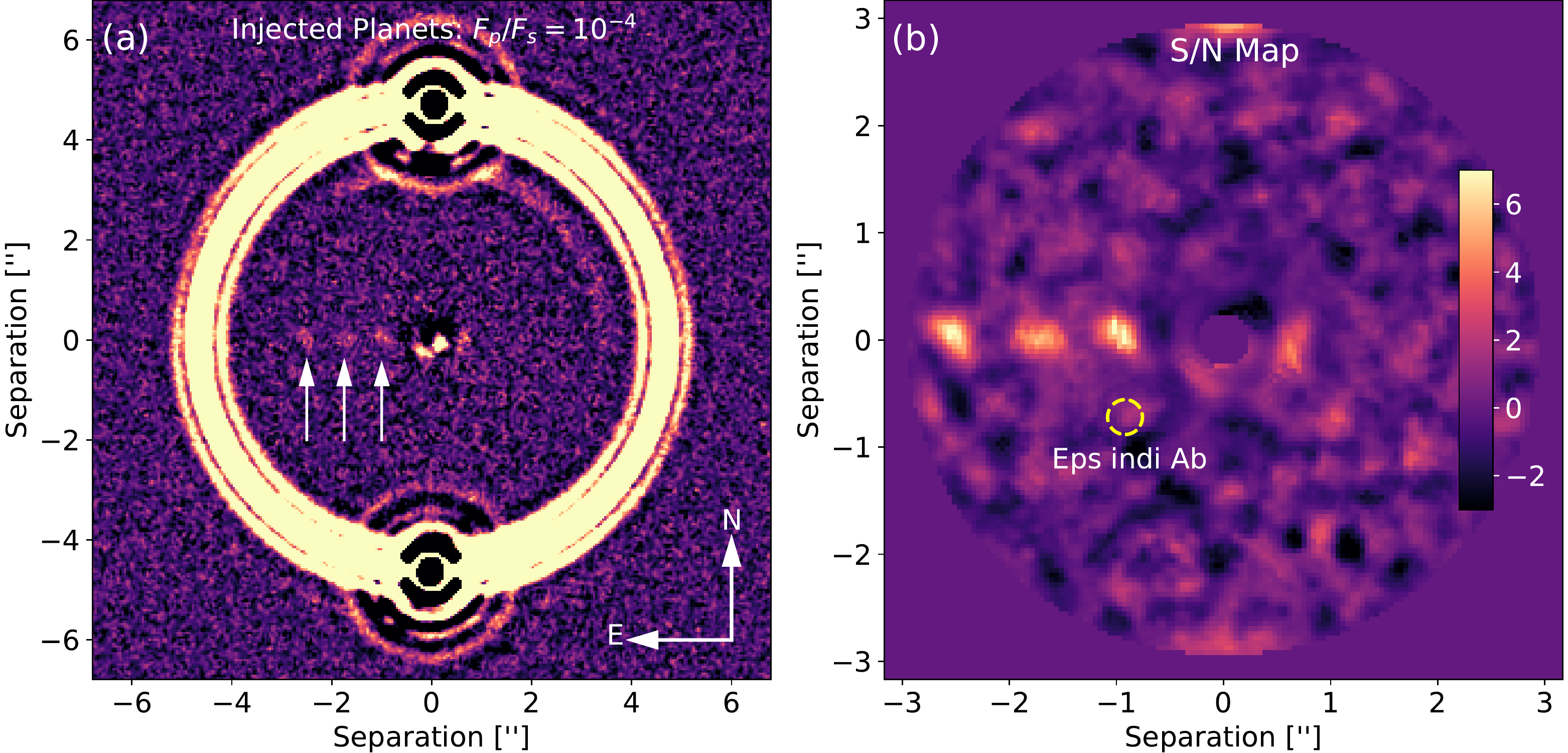}}
    \caption{\label{fig:fp_injection}(a) $\epsilon$ Indi A image with injected planets at contrast of $10^{-4}$ with a separation from $1''$ to $2.5''$ marked with white arrows to left. On the right-hand side is a likely false positive identified from examination of the odd-even frames. (b) S/N map showing injected planets appearing with S/N in the range of $5-7$. The approximate position of $\epsilon$ Indi Ab is shown by a circle, but the uncertainties are large~\citep[for details see][]{Feng2019}}.
\end{figure}

\section{Analysis}\label{sec:results}
To quantify the results further, we calculate a background noise limited sensitivity and contrast curve using artificial planet injections and recovery tests for each target, which is discussed in the following sections.  
\subsection{Background noise limited imaging performance (BLIP)}
We obtain a $5\sigma$ background noise limited imaging performance (BLIP) by calculating the standard deviation in 4-pix radius (=1.25 $\lambda$/D) apertures at various locations of the PCA reduced images. This value was then multiplied by $5$ times the square root of the number of pixels in the aperture ($5\times \sqrt{16\pi}$). To get the sensitivity with respect to the target, we used the off-axis stellar PSF for relative flux-calibration. This process was repeated for 20 angular and 17 radial distances separated by 7~pix. A mean of angular values was then calculated to get a background noise limited sensitivity curve as shown by the dashed line in the Figure~\ref{fig:combined} panel (a)-(d).     

The background noise limited sensitivity steeply increases at small separations ($\mathrm{\lesssim 1''}$) for $\epsilon$ Indi A, $\epsilon$ Eri and $\tau$ Ceti, which is due to the coronagraphic glow. Because the AGPM is not located downstream of a cold stop in VISIR, a part of the thermal emission originating from outside the telescope pupil (incl. the central obscuration) is diffracted back inside the pupil by the vortex effect (see~\citet{Absil2016} for details). This creates a significant additional amount of thermal background close to the center of the AGPM on the detector. Which could be mitigated by introducing a cold stop upstream of the AGPM. 

The BLIP sensitivity would only be reached in the absence of coronagraphic PSF residuals (quasi-static speckles --QSS--) from the central star. It can be compared to the point source sensitivity contrast introduced above to evaluate the angular separation beyond which BLIP sensitivity is reached. At such angular separations, the sensitivity improves with the square-root of integration time while sensitivity improvements in the inner regions dominated by QSS are much harder to achieve.

Figures~\ref{fig:combined} and~\ref{fig:sirius_b} show that the gap between BLIP sensitivity and point source sensitivity contrast levels out at angular separations of around 1" or 3.5 lambda/D for the fainter stars of our sample ($\epsilon$ Indi A, $\epsilon$ Eri, $\tau$ Ceti and Sirius B. At such separations, QSS are no longer seen (cf. Figure~\ref{fig:data_red}) and pixel-pixel noise dominates the sensitivity. The shallow improvement of the point source sensitivity towards even larger angular separations can be attributed to the PCA algorithm which does not conserve flux and self-subtracts a diminishing fraction of the injected fake planets signal.

The contrast around the very bright Sirius A instead is limited by QSS and PSF residuals out to an angular separation of several arcseconds. This is due to a slight misalignment between the star and the coronagraph during the observation, and to the use of a conventional Lyot-stop (LS). While the apodized LS used during the Alpha Cen observing campaign~\citep{Wagner2021} suppresses the off-axis Airy pattern at angular separations similar to the chopping throw of 4.5", this conventional LS does not reduce the Airy pattern of the off-axis chopping position and leaves residuals near the coronagraphic center which reduce sensitivity.

\subsection{Point Source Contrast Sensitivity}
To compute the planet detection sensitivity, we injected artificial planets at various angular and radial ($0.7'', 0.85'', 1'', 1.5'', 2'', 2.5''$) separations. To estimate signal-to-noise (S/N), we used the approach of~\cite{mawet2014}, as implemented in the open source Vortex Image Processing library~\citep{vip2017}. We found that a S/N of $5$ using the above criteria was sufficient to visually identify inserted artificial planets, as can be seen from Figure~\ref{fig:fp_injection}. The Figure shows injected planets at a contrast of $10^{-4}$ with respect to $\epsilon$ Indi A, with separations of $1''$, $1.7''$, $2.5''$. Panel (b) shows their S/N estimates, which varies from $5 -7$ from the inner to the outer planets.  

The overall contrast curve sensitivity was calculated by azimuthally averaging the contrast required to achieve $5~\sigma$ S/N. We used a power law function to fit the radial points to obtain final contrast curves. The contrast curves of all our targets are shown in the Figures~\ref{fig:combined} panel (a)-(d). The contrast at small separation is affected by the coronagraphic glow similar to background noise limited sensitivity, except for the target of Sirius A, for the reason mentioned in the previous section. 

Estimated contrasts of the known RV companions of $\epsilon$ Indi A and $\epsilon$ Eri are included in the Figure~\ref{fig:combined} for comparison. Planet fluxes in the NEAR filter are calculated from the ATMO 2020 spectral models \citep[][see section \ref{sec:masslims}]{atmo2020} assuming literature values for the planet mass ($3.25^{+0.39}_{-0.65}\,$\mj and $0.77\pm0.2$~\mj) and the system age ($2^{+2}_{-1.3}\,$Gyr and $0.7\pm0.3$\,Gyr). These values are plotted by blue points in the Figure~\ref{fig:combined}. Since the masses are determined by radial velocity, radius measurements are not available. Both the planet radii and temperatures are taken from the ATMO evolutionary models. 

\subsection{Mass limits}
\label{sec:masslims}
The contrast curves were converted into mass limits using the ATMO 2020 exoplanet atmosphere models \citep{atmo2020}. The models are computed using a state-of-the-art one-dimensional radiative-convective equilibrium code along a grid of self-consistent pressure–temperature profiles and chemical equilibrium abundances for a range of effective temperatures (200 to 3000 K) and gravities (2.5-5.5 dex). ATMO 2020 has key improvements over previous model families, including the use of updated molecular line opacities which results in warmer atmospheric temperature structures and improved emission spectra. We use the non-equilibrium models with weak vertical mixing and the isochrones of age-ranges obtained from literature to determine the masses corresponding to the contrast curves. 

For absolute flux value of Sirius A, we use reported value from ESO's list of mid-IR standard stars\footnote{\url{https://www.eso.org/sci/facilities/paranal/instruments/visir/tools/}}. It provides a flux of 118.8 Jy in the PAH2 filter, which has the same central wavelength (11.25 $\mu$m) as the broad NEAR filter with a bandpass of 10-12.5 $\mu$m. For the other stars, we estimated the absolute flux from the tabulated K-band magnitude and applying the K-M correction from Allen's astrophysical quantities~\citep[][]{Cox2000}. The M-N color of main sequence stars with spectral type earlier than K5 is vanishingly small as this wavelength regime is well within the Rayleigh-Jeans part of the stellar spectrum. This procedure yields the 11.25 $\mu$m fluxes of $\sim$4.5~Jy for $\epsilon$ Indi A, $\sim$7.6~Jy for $\epsilon$ Eri and $\sim$7.5~Jy for $\tau$ Ceti. We also applied the procedure to some standard stars with tabulated K-band magnitude and PAH2 fluxes and found the values to agree within a few percent. For Sirius A, the estimate is $117$~Jy in a very good agreement with the tabulated 118.8 Jy. We used Sirius A as a reference to calculate differential flux for other targets and find values agree with the calculation. We find no evidence of mid-IR excess in our targets consistent with previous work, Sirius A~\citep{white2019}, Sirius B~\citep{skemer2011}, $\epsilon$ Indi A~\citep{trilling2008}. $\epsilon$ Eri~\citep{Backman2009} and $\tau$ Ceti~\citep{lawler2014} are known to have some IR excess from their extended debris disks, but this emission is mostly from cold dust and therefore at longer wavelengths. Also, our high-spatial resolution imagery would resolve the debris disk from the central star, such that no IR excess is measured on the central PSF.

\section{Discussion}\label{sec:discussion}
\subsection{$\mathrm{\epsilon}$ Indi A}
\begin{figure*}
	\centering
    \resizebox{\textwidth}{!}{\includegraphics{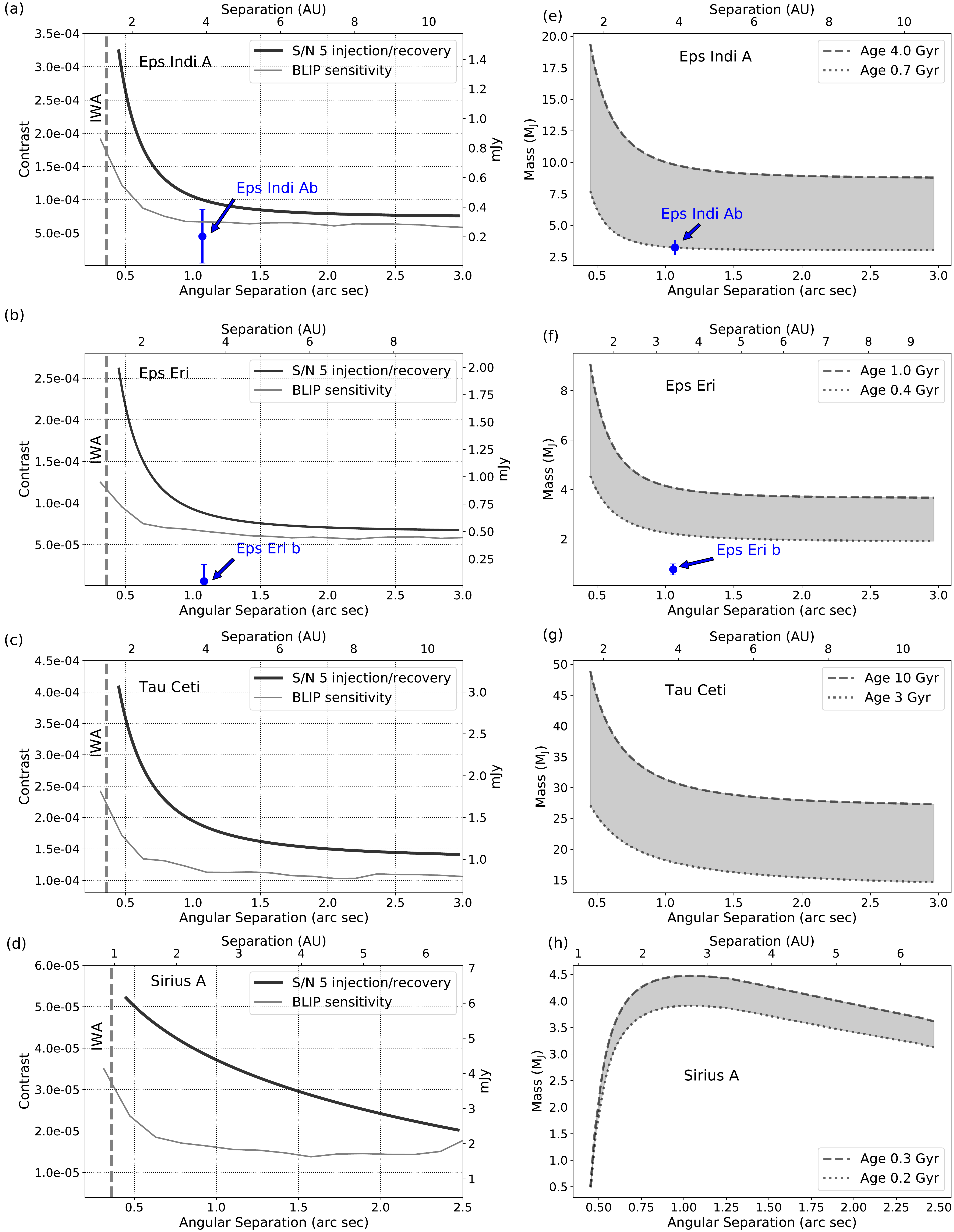}}
    \caption{\label{fig:combined} Panel (a), (b), (c), and (d) shows 5~$\sigma$ contrast curves and background noise limited imaging performance. Panel (e), (f), (g) and (h) shows the mass limits derived using contrast curves. The solid line is derived using artificial injection and recovery tests, the thin line represents the 5~$\sigma$ BLIP sensitivity. The detection limits would improve with $\sqrt{t_{obs}}$ and observing conditions. The different shape of the curves in (h) is due to the strong irradiation of the planet by the star, as discussed in Section \ref{sec:siriusA}}
\end{figure*}

To constrain the mass limits of companions to $\epsilon$ Indi A, we adopt an age range of 0.7-4\,Gyr for our models and find limits of 3.3-10\,\mj beyond 1$\arcsec$. Since the mass of the known planet is $3.25^{+0.39}_{-0.65}$,~\mj, we would have likely detected it if the age of the system was 0.7\,Gyr as shown in the Figure~\ref{fig:combined} (e). Our non-detection therefore supports an older age for $\epsilon$ Indi A. This is consistent with the majority of the age determinations as discussed in the introduction above. 
We can calculate the required observation time to detect the known giant planet by assuming a likely age of 3.8\,Gyr, in a background limited regime and improved sensitivity at small separations (without coronographic glow). It will require 50~hrs of observing time to detect the planet associated with the RV signal. 

The limits we obtain are more sensitive than any previous near IR imaging campaigns, which have constrained the masses of possible companions to 20\,\mj in the inner regions of the system \citep{Geissler2007} and 5-20\,$\mathrm{M_{J}}$ at separations larger than 2$\arcsec$ \citep{Janson2009}. Another independent reduction of $\epsilon$ Indi A combining the NEAR and NaCO L$'$ data reaches similar mass limits as ours~\citep{viswanath2021}.

$\epsilon$ Indi Ba and Bb are not discussed in this paper, as their wide orbit \citep[1459\,AU,][]{Scholz2003, McCaughrean2004} corresponds to approximately 6.7 arc minute separation, which is far outside the field of view of the VISIR instrument. 

\subsection{$\mathrm{\epsilon}$ Eri}
We adopt an age range of 0.4-1\,Gyr for our models. We obtained a mass limits of 2-4\,\mj, as shown in the Figure~\ref{fig:combined} (f). The obtained limits are more sensitive than most previous imaging data~\citep{Macintosh2003,Janson2008,Mizuki2016,Hunziker2020}, with the exception of \citet{mawet2019}, who derive an upper mass limit of about 2~\mj for an assumed system age of 400 Myr. Reaching similar mass limits at different wavelengths does, however, reduce the dependency on the planet atmosphere models. Our result therefore increases the confidence that there is indeed no planet more massive than 2~\mj around $\epsilon$ Eri, for an age of 400 Myr old. The debris disk around $\epsilon$ Eri~\citep[][]{Backman2009} is too large for our field of view and too cold and faint at $11.25~\mu$m to be detected in our observations.  
To estimate the required observing time for detecting $\epsilon$ Eri b for a likely age of 0.7\,Gyr, we find that 1~\mj planet can be detected in less than 70~hrs with the current setup and sensitivity will improve at small separations, if the coronographic glow is eliminated (by introducing a cold pupil stop in front of the AGPM coronagraph mask).

\subsection{$\tau$ Ceti}
Because of the large range of age suggested in the literature, we adopt a wide range of 3-10\,Gyr. The obtained mass limits of 15-30$\,$\mj shown in the Figure~\ref{fig:combined} (g) are comparable to those of \citet{boehle2019}, who use a combination of radial velocity and $3.8\,\mathrm{\mu}$m imaging data to find limits of 10-20~\mj beyond 2$\arcsec$ for a slightly younger age range (2.9-8.7\,Gyr). Within 2$\arcsec$ we are more sensitive, and the non-detection of any of the known planets is consistent with the expectation, as the various Earth-mass planets of the system are far too low mass and are located within the inner working angle of our data and even the proposed giant planet candidate is well below our detection threshold (1-2~\mj at 3-20\,AU).

\subsection{Sirius A}
\label{sec:siriusA}
For our models we assumed an age range of 0.2-0.3\,Gyr for Sirius. For Sirius A we have also included the irradiation from the star when determining the expected magnitude of the planet, due to the brightness of the host star. We accounted for the irradiation by counting both the flux arriving at the planet surface and the intrinsic formation heat of the planet from the model as incoming energy when calculating the equilibrium temperature, which in turn determines the outgoing flux. Sirius A is 25 times more luminous than the Sun, and a potential planet on a 2 AU orbit would be heated to more than 400 K independent of the planet's age or mass. This leads to the peculiar shape of the mass contrast curve with better mass sensitivities at small separation from Sirius A shown in the Figure~\ref{fig:combined} panel h. As a result, the curve of the mass limit falls of sharply within $\sim$1$\arcsec$ or 2.6 AU, where the stellar radiation dominates. Beyond this the irradiation quickly becomes negligible due to falling off with the square of the distance and the curve looks more similar to those of the other systems. The other systems have smaller, fainter stars and therefore the amount of heating by the stellar flux is negligible at the separations resolved by our imagery. While we do not detect any planets, we do obtain the most sensitive mass limits to date within 1.5$\arcsec$ and comparable limits to the most sensitive limits of previous imaging campaigns outside that radius \citep{Bonnet2008, Thalmann2011, vigan2015, Hunziker2020}. We cannot rule out the possibility of planets that could be hidden behind or too close to the star. 

\subsection{Sirius B}
\begin{figure*}
	\centering
    \resizebox{\hsize}{!}{\includegraphics{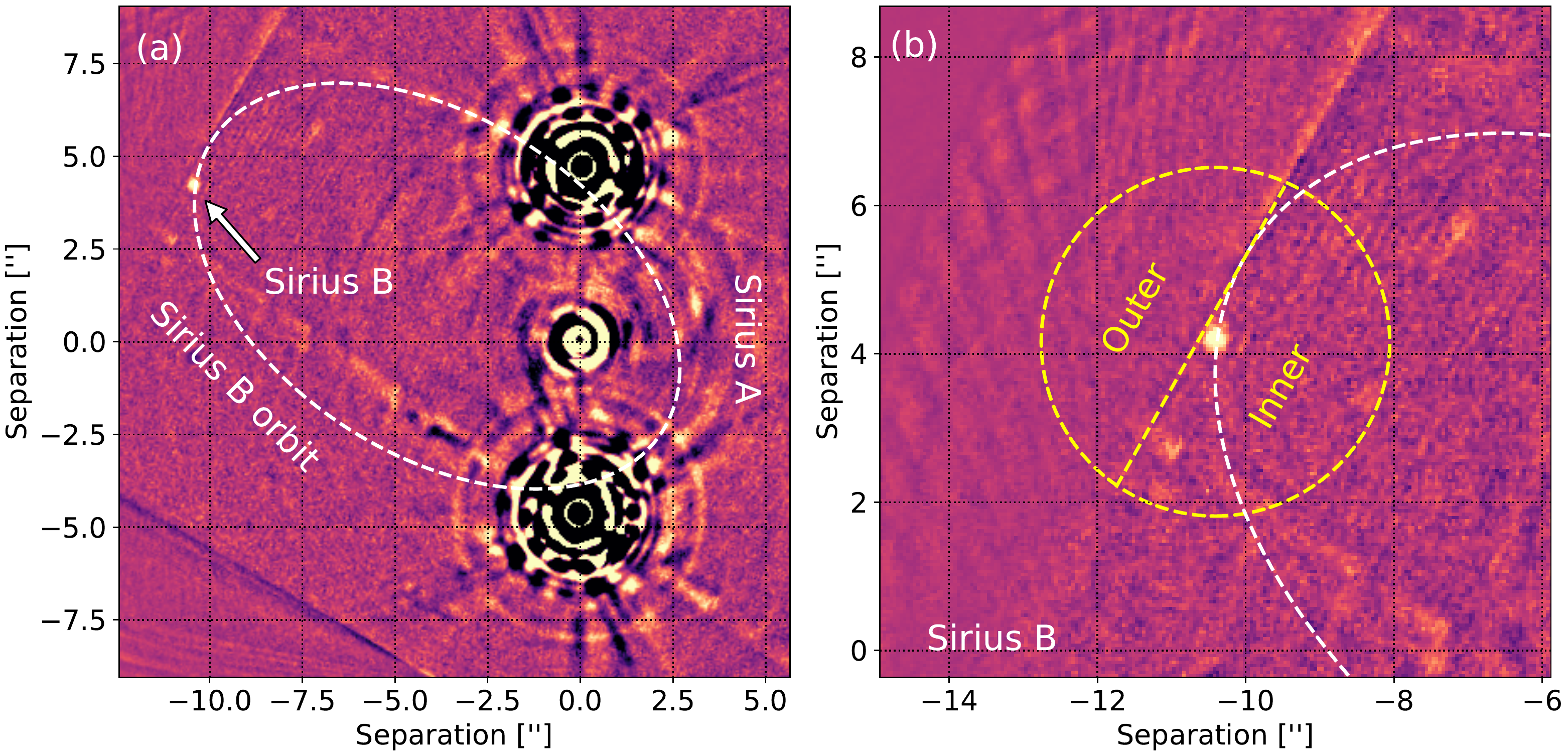}}
    \caption{\label{fig:sirius_b} (a) Final derotated image showing Sirius A and Sirius B. (b) Zoomed in on Sirius B, divided into two regions (inner and outer) based on the number of frames in the outlined area.}
\end{figure*}

No ADI reduction had to be applied to the Sirius B data, because the star is $4\times10^{-5}$ times fainter than Sirius A and no PSF residuals are seen besides the PSF core. Also the target was far from Sirius A and the noise of its coronagraphic PSF, so no further processing was necessary. The final image of Sirius A and B is shown in the Figure~\ref{fig:sirius_b}. Sirius B was very close to edge of the NEAR's field of view, and even outside of it for some individual frames, which were then excluded before derotation and averaging. To calculate the sensitivity around Sirius B, we divided that area of the image into two regions, an inner and an outer region, as shown in the Figure~\ref{fig:sirius_b} (b). The inner region provides a better sensitivity for both the background noise limited and injected planet, as it includes more frames.  

The astrometry measurements for Sirius B, put it at a separation of $11.18''$ from Sirius A. Relative photometry with respect to Sirius A has been performed using an optimum r=4~pix photometric aperture. The $11.25~\mu$m contrast ratio between the two stars is $4\times10^{-5}$ corresponding to a Sirius B flux of 4.7 mJy at an S/N of about 40. This value is consistent with the low S/N measurement of 4.9 mJy reported for a similar observing band by~\citet{skemer2011}.

For our models we have assumed the same age range as we did for Sirius A, giving us limits of $1.5-1.8$\,\mj for the "inner" region and $3.1-3.6$\,\mj for the "outer" region. The mass limits in the inner region are comparable to previous near IR limits of 1.6\,\mj from~\citet{Thalmann2011}. According to the models and assuming the sensitivity improving with $\sqrt{t_{obs}}$, a $4\times$ more observation time would be enough to reach 0.5~\mj.

\begin{figure}
	\centering
    \resizebox{\hsize}{!}{\includegraphics{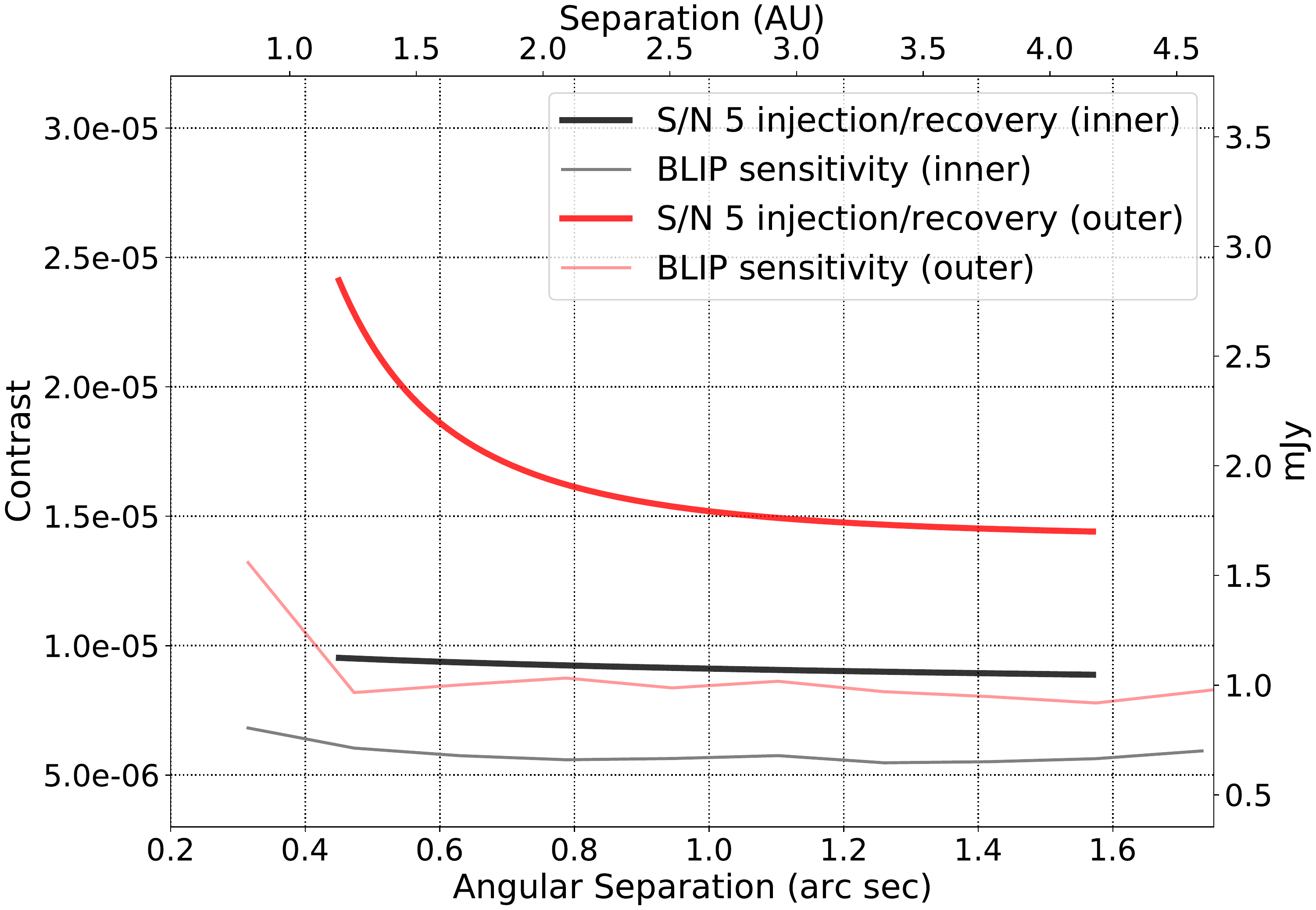}}
    \caption{\label{fig:siriusB_cc} Sirius B 5~$\sigma$ contrast curves, same as in Figure~\ref{fig:combined}. The outer and inner regions are represented by red and black lines respectively. The contrast curve for outer region increases at small separation due to edge of the inner frames.}
\end{figure}

\begin{figure}
	\centering
    \resizebox{\hsize}{!}{\includegraphics{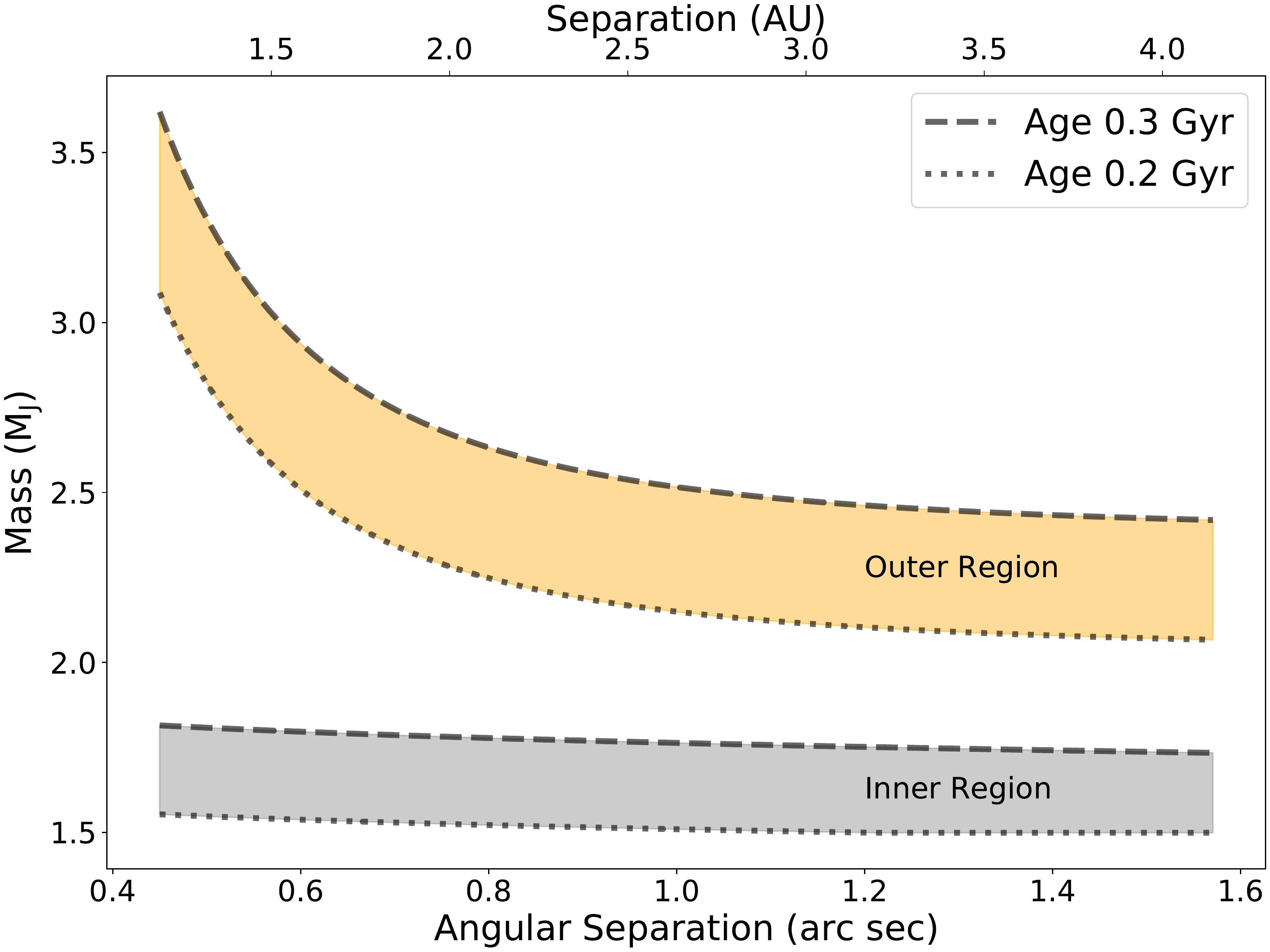}}
    \caption{\label{fig:mcurve_siriusB} Planet mass upper limits derived for Sirus B. The two mass limits are based on the outer and inner region, see~Figure~\ref{fig:siriusB_cc}.}
\end{figure}

\section{Summary and Conclusions}\label{sec:conclusion}
In this work we demonstrate high-contrast imaging of a small sample of very nearby (<4 pc) stars with spectral type earlier than M at $11.25~\mu m$ with NEAR. While we do not detect any known or new planets, we are able to set upper mass limits of the order of a few Jupiter masses for most of the targets and of $15-30$\,\mj for the older $\tau$ Ceti. For $\epsilon$ Indi A and $\epsilon$ Eri we achieve detection limits very close to the giant planets discovered by RV, with the limits on $\epsilon$ Indi A being the most sensitive to date. Also for $\tau$ Ceti and Sirius A we obtain the most sensitive limits to date at small separations (<1.5$\arcsec$ and <2$\arcsec$ respectively). Our mass limit for Sirius B is similar to the one achieved previously at a shorter wavelength.

The ADI analysis was performed using a PCA based algorithm with artificial planet injections and recovery tests yielding some of the most stringent upper mass limits to date. For $\epsilon$ Indi A and $\epsilon$ Eri, we almost reach the detection limit for the known planets if they were at the young end of the possible age estimates but fail to detect any signal. We achieve an unprecedented sub-mJy detection sensitivity.

We demonstrate close to background noise limited imaging for most of our target stars apart from the glaring Sirius A, for which the data is contrast limited. Assuming likely ages and background noise limited imaging, the giant planets orbiting $\epsilon$ Eri and $\epsilon$ Indi A can be imaged in 70 and 50~hrs respectively. Finally, for our closest and youngest (together with Sirius A) target Sirius B, the sub-Jupiter mass regime could be reached by merely doubling the observation time.

This work shows the potential of direct imaging in the mid-IR regime and prospects for upcoming mid-IR HCI instruments. Upcoming mid-IR HCI instrument such as METIS at the ELT would be able to detect known planets around $\epsilon$ Indi A and $\epsilon$ Eri in a few minutes of observation time and reach sensitivities to detect Earth size exoplanets in a few hours~\citep[][]{brandl2018}.

\begin{acknowledgements}
The authors would like to thank the ESO and the Breakthrough Foundation and all the people involved for making the NEAR project possible. The observations were carried out under the ESO program id: 60.A-9107(D) and  60.A-9107(F). MRM acknowledges the support of a grant from the Templeton World Charity Foundation, Inc. The opinions expressed in this publication are those of the authors and do not necessarily reflect the views of the Templeton World Charity Foundation, Inc. Part of this work has received funding from the European Research Council (ERC) under the European Union's Horizon 2020 research and innovation programme (grant agreement No. 819155), and by the Wallonia-Brussels Federation (grant for Concerted Research Actions).
\end{acknowledgements}
 
\bibliographystyle{aa}
\bibliography{bibliography}

\end{document}